\documentclass[conference,a4paper]{IEEEtran}
\usepackage{xcolor}
\usepackage{balance}
\usepackage{hhline}
\usepackage{textgreek}
\usepackage{textcomp}
\usepackage{gensymb}
\usepackage{amssymb}
\usepackage{soul}
\usepackage{multirow}

\ifCLASSINFOpdf
  \usepackage[pdftex]{graphicx}
\else
  \usepackage[dvips]{graphicx}
\fi
\usepackage{amsmath}

\usepackage{subcaption}
\begin{document}
%

\title{Measurement and Characterization of an Indoor Industrial Environment at 3.7 and 28 GHz}


\author{\IEEEauthorblockN{
Mathis Schmieder\IEEEauthorrefmark{1},   
Taro Eichler\IEEEauthorrefmark{2},   
Sven Wittig\IEEEauthorrefmark{1},    
Michael Peter\IEEEauthorrefmark{1},      
Wilhelm Keusgen\IEEEauthorrefmark{1}
}                                     
\IEEEauthorblockA{\IEEEauthorrefmark{1}
Fraunhofer Heinrich Hertz Institute, Berlin, Germany, mathis.schmieder@hhi.fraunhofer.de}
\IEEEauthorblockA{\IEEEauthorrefmark{2}
Rohde \& Schwarz, Munich, Germany}
}

\maketitle

\begin{abstract}
Fifth generation (5G) mobile networks are expected to play an increasing role in industrial communication with private mobile communication networks deployed on company premises. For planning, standardization and product development, it is crucial to to thoroughly understand the radio channel characteristics of such environments. Frequencies around 3.7\,GHz were already reserved by regulation authorities and to meet the increasing demand for higher bandwidths, spectrum in the millimeter wave range around 28\,GHz is targeted.
This paper presents a wideband channel measurement campaign at both 3.7 and 28\,GHz with direction-of-arrival information at 28\,GHz. The results are compared to the 3GPP TR 38.901 Indoor Factory model and to two other recent papers. Evaluation of path loss and RMS delay and angle spread show the unique nature of industrial indoor environments. 
\end{abstract}

\vskip0.5\baselineskip

%

\vspace{-2pt}
\section{Introduction}

The integration of new digital technologies for intelligent manufacturing, known as Industry 4.0, is about to substantially transform industrial production \cite{draht2014industrie4}.  The technologies allow all data of the production process to be analyzed and exploited, making the process faster, more flexible and more cost-effective. Industry 4.0 will change traditional relationships among suppliers, producers, and customers as well between human and machine. A key element for these developments is wireless communication between machines, robots and humans – depending on the application, with both low latency and high throughput. Fifth generation (5G) mobile network technologies are supposed to meet these requirements \cite{Wollschlaeger2017}. For reasons of data sovereignty, companies are increasingly interested in operating their own mobile communications network on their premises. Initially, such networks will be implemented at frequencies below 6 GHz. In Germany, for example, the regulation authority has reserved the frequency range from 3.7 to 3.8 GHz for this purpose. Since it is foreseeable that the demand for bandwidth will further increase, spectrum in the millimeter wave range between 24 and 26 GHz is already targeted for use in the industrial context.
To evaluate the performance of such communication systems during standardization and product development, it is crucial to use a channel model suitable for the propagation environment, including appropriate model parameters. Geometry-based stochastic channel models (GSCMs) have become widely adopted for this purpose. The 3rd Generation Partnership Project (3GPP) has proposed models for typical mobile frequencies around 2 GHz, but the latest model 38.901 Version 16.0.0 \cite{3GPP-TR-38901} even targets the range from 0.5 to 100 GHz. Very recently it has been extended to cover also the industrial environments with a scenario called \emph{Indoor Factory}. Such channel models rely heavily on measurement data obtained in typical propagation environments. In \cite{schmieder2019directional}, the authors of this paper summarized several measurement campaigns mostly at sub-6\,GHz frequencies and presented first angle-resolved measurement results at 28\,GHz in an industrial environment. Another recent paper by Jaeckel et al. \cite{jaeckel2019industrial} describes industrial indoor measurements at several scenarios from 2 to 6\,GHz.

This paper presents a wideband channel measurement campaign at 3.7 and 28\,GHz in an industrial environment. Power delay profiles, path loss and root mean square (RMS) delay spreads are evaluated for 3.7 and 28\,GHz. Additionally, direction-of-arrival (DoA) is extracted and the RMS angular spread is evaluated for 28\,GHz. The results are compared to the recently standardized \emph{Indoor Factory} scenario described in 3GPP TR 38.901 \cite{3GPP-TR-38901}. Additionally, the results at 3.7\,GHz are compared to the findings of Jaeckel et al. \cite{jaeckel2019industrial} and at 28\,GHz to recent results of the authors of this paper \cite{schmieder2019directional} that were conducted using the same channel sounder setup.

\vspace{-2pt}
\section{Measurement Scenario and Procedure}
\begin{figure*}[htb]
\vspace{.1em}
\centering
\includegraphics[width=\textwidth]{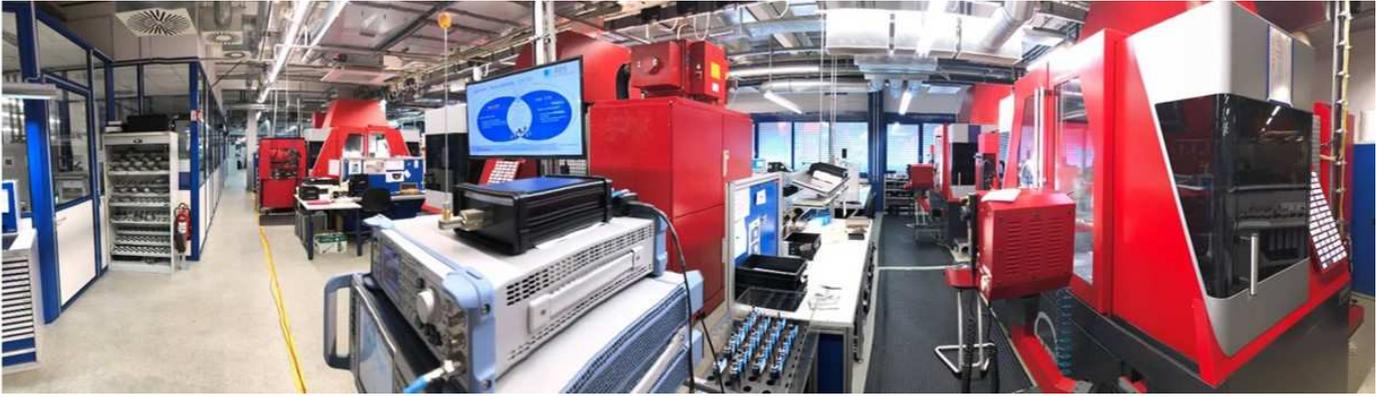}
    \caption{Overview of the measurement scenario from BS position}
    \vspace{-1em}
    \label{fig:scenario}
\end{figure*}

\begin{figure}[htb]
    \centering
    \includegraphics[width=0.45\textwidth]{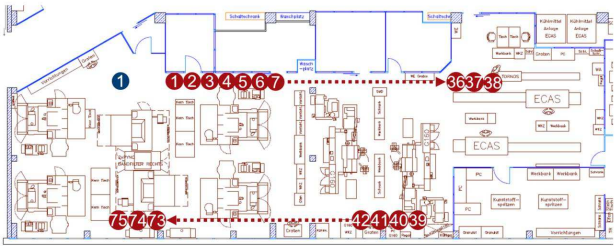}
    \caption{Schematic of the factory floor space}
    \label{fig:floorplan}
\end{figure}

The measurements were conducted inside a high-precision machining workshop hall with a maximum length of 40.5 meters and maximum width of 15.5 meters and a total floor space area of 464 square meters. The height of the ceiling is 4.6 meters and the detailed schematic of the floor space is shown in Fig. \ref{fig:floorplan}. The factory floor is densely packed with various industrial machines (CNC milling machines etc.), to a large extent consisting of metal or with metal surfaces. Enclosing the hall are glass windows on one side and walls (partly glass) to neighboring conference rooms and other various purpose rooms on the opposite side.
Fig. \ref{fig:scenario} shows the factory hall scenario from the base station perspective (transmitter).

The measurements were conducted in the following way: The transmitter (Tx) was placed at a “base station” (BS) position at a height of 185 cm and was fixed for all measurements while the receiver (Rx), emulating user equipment (UE) was positioned at 75 different positions on the floor (as depicted in Fig. \ref{fig:floorplan}). The height of the receiver was 144 cm. The positions 1-38 were line-of-sight (LOS), while the positions 39-75 were non-line-of-sight (NLOS).

\vspace{-2pt}
\section{Channel Sounder Setup}
\begin{figure}[htb]
    \centering
    \includegraphics[width=0.45\textwidth]{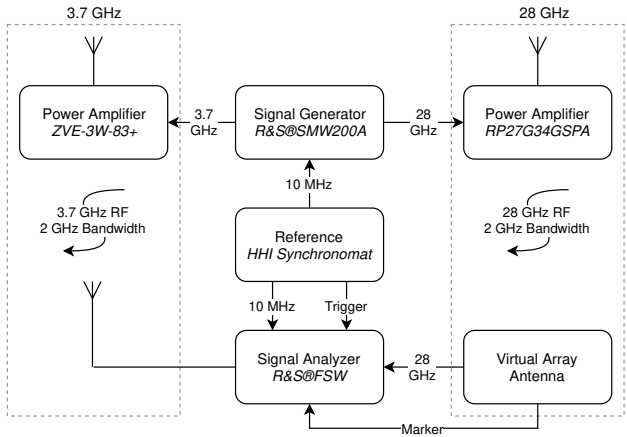}
    \caption{Channel sounder setup}
    \label{fig:setup}
\end{figure}

The measurements were conducted using an instrument-based highly flexible time-domain channel sounder as described in \cite{schmieder2019directional}. The block diagram in Figure \ref{fig:setup} shows a simplified block diagram of the used setup.

The transmitter (Tx) comprises a vector signal generator (R\&S\textsuperscript{\textregistered}SMW200A), a power amplifier (PA) and a vertically polarized omni-directional $\lambda/4$ ground plane antenna for each used frequency band. At both frequencies, a periodic 96,000 sample Frank-Zadoff-Chu (FZC) \cite{chu1972polyphase}, \cite{frank1973comments} with a duration of 48\,\textmu s and a bandwidth of 2\,GHz was used as channel sounding signal and generated directly at the RF frequencies. By using external up and down converters, the channel sounder setup can also be used at V, E and W bands. The total transmit powers were 30\,dBm at 3.7 and 34\,dBm at 28\,GHz.

The receiver (Rx) comprises a receive antenna and a signal analyzer (R\&S\textsuperscript{\textregistered}FSW) that captures the signal and streams the baseband samples with a resolution of 16\,bit to a connected Laptop. At 3.7\,GHz, a similar $\lambda/4$ ground plane antenna as for the transmitter was used. At 28\,GHz, the antenna was a virtual circular array antenna (VCA) \cite{nguyen2016instantaneous} similar to the setup in \cite{schmieder2019directional}. Both transmitter and receiver shared a common clock provided by a high precision rubidium clock (Synchronomat by \emph{Fraunhofer HHI}), that also enabled coherent triggering at the receiver. The important channel sounder parameters are summarized in Table \ref{table_sounderparameters}.

\begin{table}[htb]
\renewcommand{\arraystretch}{1.0}
\caption{Channel sounder parameters}
\label{table_sounderparameters}
\centering
\begin{tabular}{|c|c|c|}
\hline
Parameter & \multicolumn{2}{c|}{Value}\\
\hhline{|=|=|=|}
Carrier frequency & 3.7 GHz & 28 GHz\\
\hline
Transmit power & 30 dBm & 34 dBm \\
\hline
Sounding bandwidth & \multicolumn{2}{c|}{2000 MHz} \\
\hline
Sampling rate at Tx & \multicolumn{2}{c|}{2400 MHz} \\
\hline
Sampling rate at Rx & \multicolumn{2}{c|}{2500 MHz} \\
\hline
Sequence duration & \multicolumn{2}{c|}{48 \textmu s} \\
\hline
Temporal snapshot separation & \multicolumn{2}{c|}{48 \textmu s} \\
\hline
Instantaneous dynamic range & \multicolumn{2}{c|}{65\,dB} \\
\hline
Diameter of virtual array & N/A & 97.8 mm \\
\hline
\# of virtual array elements & N/A & 1000 \\
\hline
Distance between elements & N/A & 0.307 mm (0.0287 $\lambda$) \\
\hline
\end{tabular}
\end{table}

\vspace{-2pt}
\section{Measurement Results and Evaluation}
For every measurement point, the channel sounder generated 1000 channel impulse response (CIR) snapshots. At 3.7\,GHz, the 1000 snapshots were coherently averaged and calculated into instantaneous power delay profiles (IPDP). At 28\,GHz, each of the CIR snapshot corresponds to one virtual array antenna element. By incoherently averaging the CIR snapshots into average power delay profiles (APDP), the variance in power can be greatly reduced, resulting in a much smoother gradient of the APDP. Prior to further estimation of large scale parameters and, at 28\,GHz, direction-of-arrival (DoA) information, all components before the theoretical LOS delay were removed from both the IPDPs and APDPs.

\begin{figure}[htb]
\vspace{.1em}
\begin{subfigure}{.48\textwidth}
    \centering
    \includegraphics[width=0.95\textwidth]{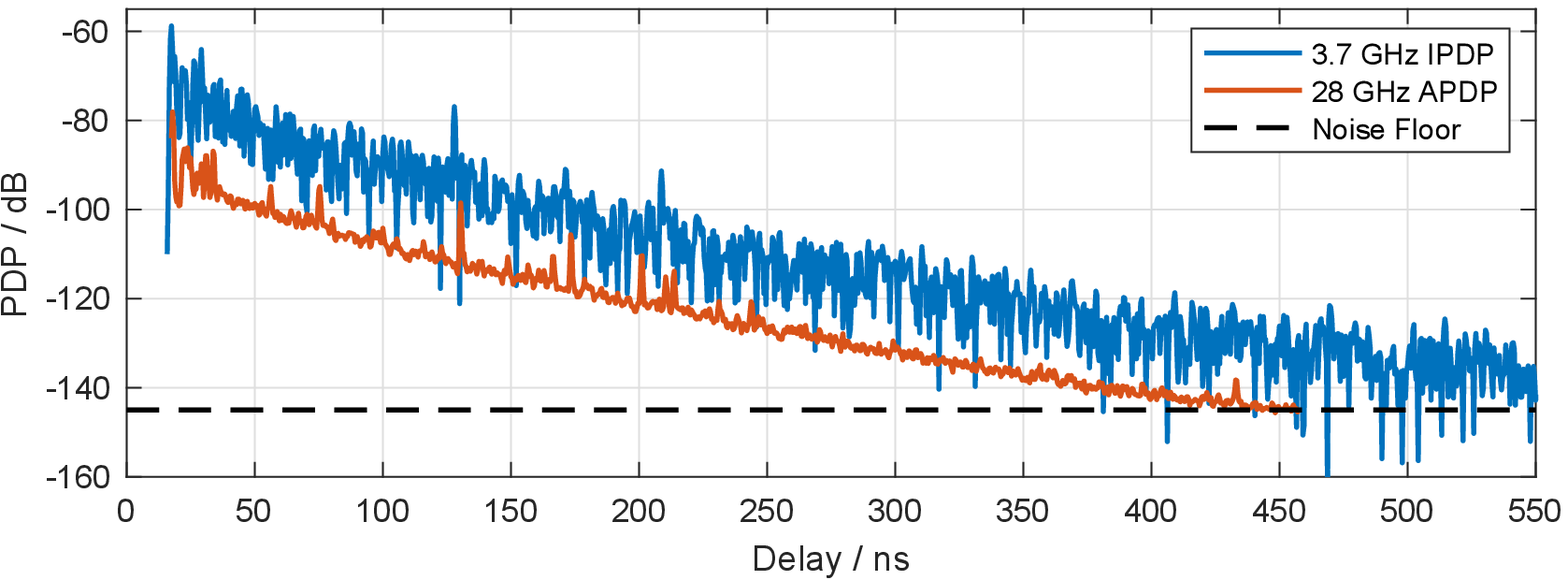}
    \caption{LOS}
    \label{fig:pdp_los}
\end{subfigure}

\begin{subfigure}{.48\textwidth}
    \centering
    \includegraphics[width=0.95\textwidth]{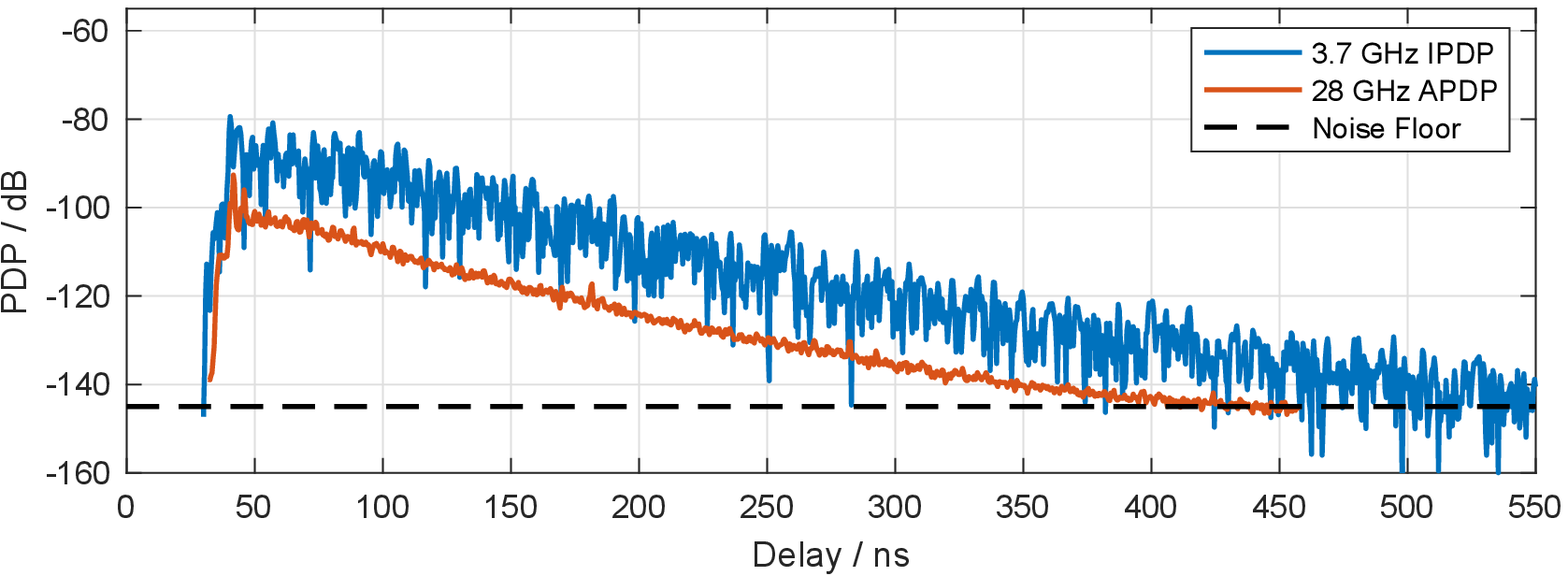}
    \caption{NLOS}
    \label{fig:pdp_nlos}
\end{subfigure}
\caption{IPDP and APDP in LOS and NLOS condition}
\label{fig:pdp}
\end{figure}

Figure \ref{fig:pdp} shows IPDPs at 3.7\,GHz and APDPs at 28\,GHz in both LOS (\ref{fig:pdp_los}) and NLOS (\ref{fig:pdp_nlos}) condition. A higher variance in power at 3.7\,GHz compared to 28\,GHz can clearly be seen. The LOS power delay profiles (PDP) correspond to a measurement point with a Tx-Rx-distance of 5.2 meters, resulting in a time of flight delay of 17.5\,ns. At 3.7\,GHz, the LOS component was received with a power of -58.77\,dBm and at 28\,GHz with a power of -78.08\,dBm. For both frequencies, several specular multipath components (MPC) can be seen together with dense multipath components (DMC) \cite{putanen_dmc} that fall below the noise floor of -145\,dBm at a delay of 457\,ns at 28\,GHz and 550\,ns at 3.7\,GHz.

The NLOS PDPs correspond to a measurement point with a Tx-Rx-distance of 9.9 meters, resulting in a time of flight delay of 33\,ns. Only few strong MPCs can be seen at 3.7\,GHz with a delay of 40\,ns and a power of -79.41\,dBm. The DMC fall below the noise floor at a delay of 455\,ns at 28\,GHz and at 550\,ns at 3.7\,GHz.

In order to estimate large scale parameters and DoA information, an evaluation threshold of 40\,dB relative to the strongest component was applied to the PDPs. Local maxima were identified in the PDPs and used to evaluate path loss and delay spread. The CLEAN technique \cite{tsao1988reduction} with real-valued beamspace MUSIC (RB-MUSIC) \cite{mathews1994eigenstructure} as a DoA estimator was applied to the CIR snapshots at 28\,GHz in order to extract DoA information. Based on this information, the angular spread and angular power spectrum were estimated. All evaluations were done separately for LOS and NLOS.

\vspace{-2pt}
\subsection{Path Loss}
\begin{figure}[htb]
\begin{subfigure}{.48\textwidth}
    \centering
    \includegraphics[width=0.95\textwidth]{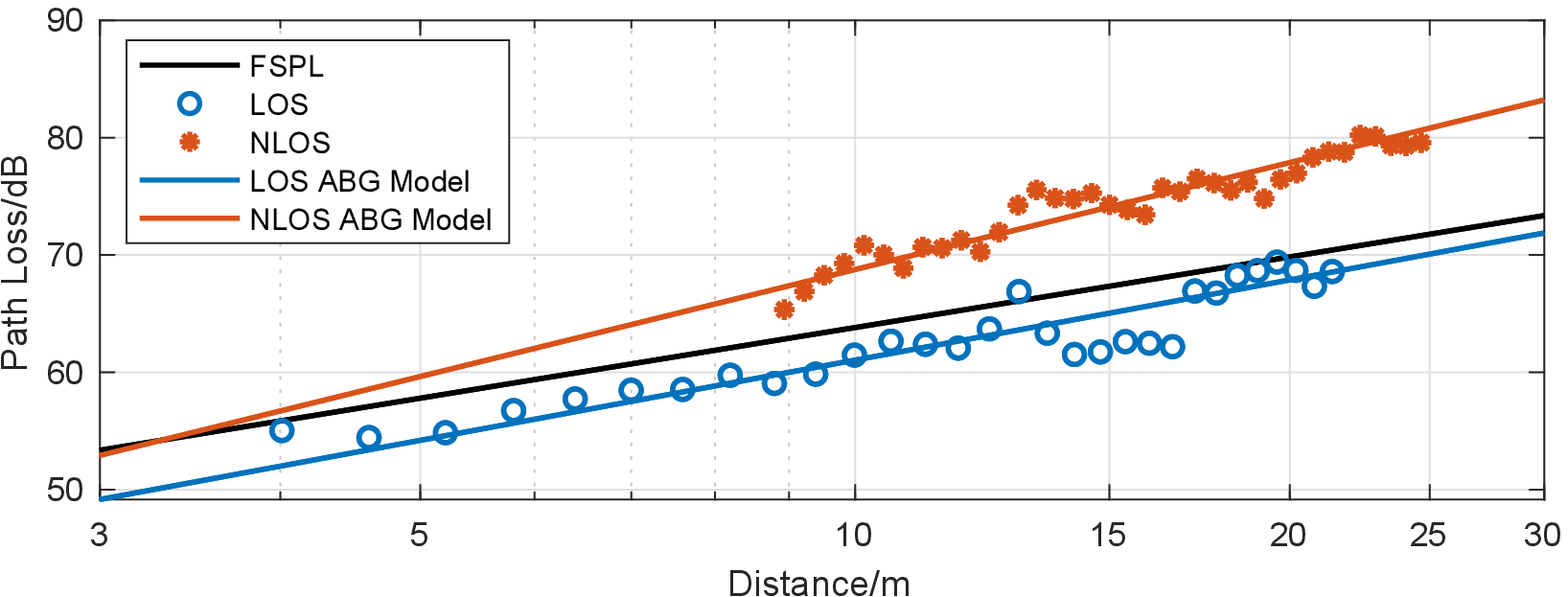}
    \caption{3.7\,GHz}
    \label{fig:pathloss_37}
\end{subfigure}

\begin{subfigure}{.48\textwidth}
    \centering
    \includegraphics[width=0.95\textwidth]{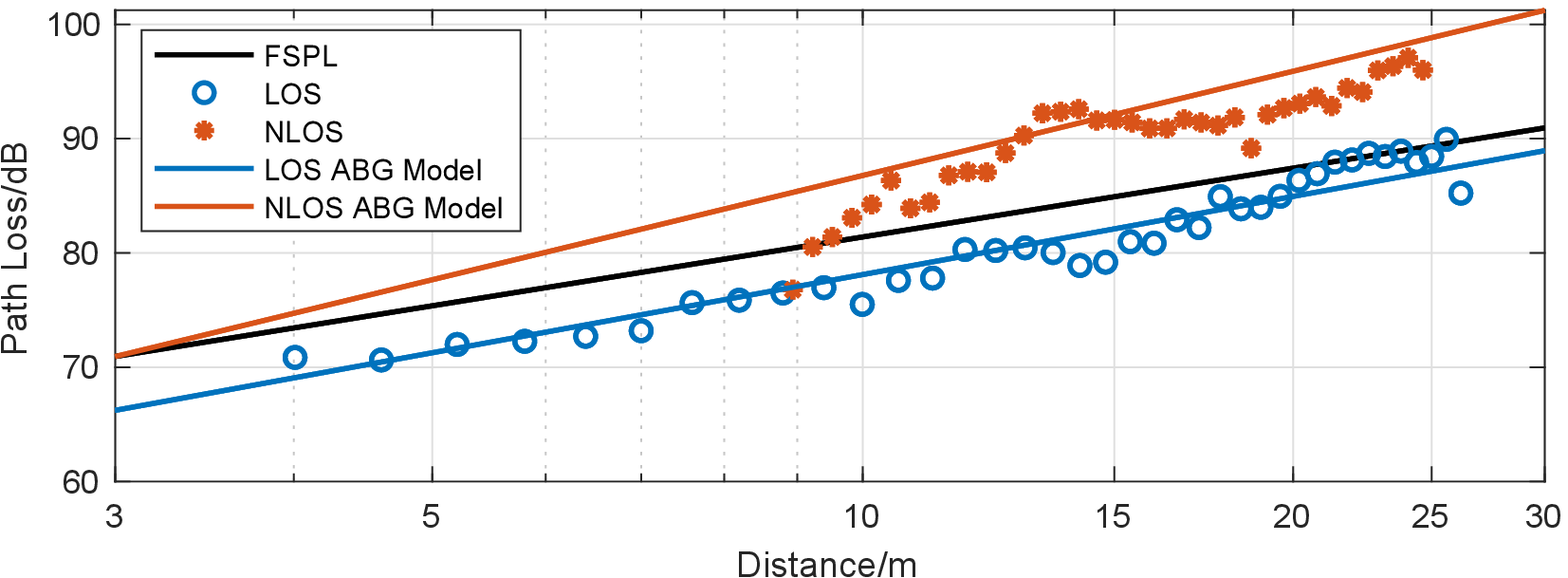}
    \caption{28\,GHz}
    \label{fig:pathloss_28}
\end{subfigure}
\caption{LOS/NLOS Path Loss}
\label{fig:pathloss}
\end{figure}

The path loss was evaluated based on the accumulated powers of the identified local maxima in the IPDPs and APDPs. In Figure \ref{fig:pathloss}, the results of the evaluation are illustrated for 3.7 (\ref{fig:pathloss_37}) and 28\,GHz (\ref{fig:pathloss_28}). Data corresponding to LOS condition is coloured blue, while NLOS data is coloured red. The theoretical free space path loss (FSPL) is added in black. In LOS condition, the minimal distance between Tx and Rx was 4 meters and the maximum distance was 26 meters, in NLOS conditions the measurements were conducted between 9 and 25 meters Tx-Rx distance.

At both frequencies, the LOS path loss for distances of up to 20 meters is about 3\,dB less than the FSPL but grows closer to FSPL for higher distances. The NLOS path loss at distances around 10 meters is close to FSPL and grows with increasing distance. At 10 meters, the NLOS path loss is about 6\,dB higher than FSPL at 28\,GHz and almost 10\,dB higher than FSPL at 3.7\,GHz.

For both LOS and NLOS, ABG path loss model \cite{sun2016propagation} parameters according to (\ref{eq:pathloss_abg}) were derived using least squares fitting. The coefficients $\alpha$ and $\gamma$ show the dependence of path loss on 3D Rx-Tx distance $d$ and frequency $f$ relative to $d_0 = 1\,\textrm{m}$ and $f_0 = 1\,\textrm{GHz}$ and $\beta$ is an optimized offset value for path loss in dB. $\chi^{\mathrm{ABG}}_\sigma$ is the lognormal random shadowing variable with corresponding standard deviation $\sigma$. Table \ref{tab:pathloss} shows the derived path loss model parameters.

\begin{equation}
	\begin{split}
		PL_{\mathrm{ABG}}(f,d) &= 10 \alpha\,\log_{10} \left( \frac{d}{d_0} \right) + \beta \\
							   &+ 10 \gamma\,\log_{10} \left( \frac{f}{f_0} \right) + \chi^{\mathrm{ABG}}_\sigma
	\end{split}
	\label{eq:pathloss_abg}
\end{equation}

Analysis of the derived LOS model shows that, while it is a good fit to the measurement results between 5 and 26 meters Rx-Tx distance, the model produces unrealistic low path loss values for distances shorter than 5 meters. Similarly, the NLOS model is not a good fit for the measurement results at distances lower than 10 meters, especially at 28\,GHz. In this environment, dual slope models with a break point around 10 meters distance would have been a better fit. More NLOS measurements for lower distances would be necessary to thoroughly parameterize such model.

\begin{table}[htb]
\renewcommand{\arraystretch}{1.0}
\vspace{.5em}
\caption{Path loss model parameters for $d_0=1\,\mathrm{m}$}
\label{tab:pathloss}
\centering
\begin{tabular}{|c|c|c|c|c|}
\hline
 & PLE / $\alpha$ & $\beta$ & $\gamma$ & $\sigma$ \\
 &                & (dB)    &          &  (dB)    \\
\hline
LOS       & 2.27 & 27.29 & 1.94 & 1.62  \\
\hline
NLOS      & 3.02 & 28.35 & 1.78 & 1.61  \\
\hline
3GPP LOS  & 2.15 & 31.84 & 1.90 & 4.30 \\
\hline
3GPP NLOS & 3.57 & 18.60 & 2.00 & 7.20 \\
\hline
Jaeckel et al. \cite{jaeckel2019industrial} LOS & 1.83 & 36.30 & 1.95 & 1.63 \\
\hline
Jaeckel et al. \cite{jaeckel2019industrial} NLOS & 2.41 & 29.10 & 2.54 & 3.58 \\
\hline
\end{tabular}
\end{table}

Comparison of the parameters with the 3GPP TR 38.901 \cite{3GPP-TR-38901} \emph{Indoor Factory} model in the \emph{Dense, Low} scenario show that, while the LOS parameters are of similar magnitude, the 3GPP model does not agree well with the scenario described in this paper. Similarly, the results of Jaeckel et al. \cite{jaeckel2019industrial}, which are valid from 2 to 6\,GHz, do not agree well with the parameters in this paper. In order to compare the path loss model to the results in \cite{schmieder2019directional} at 28\,GHz, the ABG model parameters have to be converted to a floating intercept (FI) model according to \ref{eq:pl_abg_fi}

\begin{equation}
        PL^{\mathrm{FI}}_{0} = \beta + 10 \gamma\,\log_{10} \left( \frac{f}{f_0} \right)
    \label{eq:pl_abg_fi}
\end{equation}

with $n^{\mathrm{FI}} = \alpha$ resulting in $PL^{\mathrm{FI}}_{0,\mathrm{LOS}} = 55.36\,\textrm{dB}, n^{\mathrm{FI}}_{\mathrm{LOS}} = 2.27$ and $PL^{\mathrm{FI}}_{0,\mathrm{NLOS}} = 54.11\,\textrm{dB}, n^{\mathrm{FI}}_{\mathrm{NLOS}} = 3.02$. Compared to the findings in \cite{schmieder2019directional} with $PL_{0,\mathrm{LOS}}(d_0)= 59.7\,\textrm{dB}$, $n_{\mathrm{LOS}} = 1.8$ and $PL_{0,\mathrm{NLOS}}(d_0)= 77.4\,\textrm{dB}$, $n_{\mathrm{NLOS}} = 0.9$ show that the model is not a good fit for the measurement results in this paper. This highlights that, at least in industrial environments, the path loss model is highly scenario specific.

\vspace{-2pt}
\subsection{RMS Delay Spread}
After applying an evaluation threshold of 40\,dB relative to the strongest component to the IPDPs and APDPs, the RMS delay spread (DS) was calculated for both LOS and NLOS. Table \ref{tab:delayspread} shows the statistical parameters: mean $\mu_{\mathrm{DS}}$, median $m_{\mathrm{DS}}$, standard deviation $\sigma_{\mathrm{DS}}$ and 95\%-quantile $Q_{\mathrm{DS},95}$. In order to compare the parameters to the 3GPP \emph{Indoor Factory} \cite{3GPP-TR-38901} model, the parameters were also evaluated on a logarithmic scale.

\begin{table}[ht]
\renewcommand{\arraystretch}{1.0}
\caption{Statistical parameters of the RMS DS with 3GPP InF \cite{3GPP-TR-38901} for comparison}
\label{tab:delayspread}
\centering
\begin{tabular}{|c|c|c|c|c|c|}
\hline
 & & $\mu_{\mathrm{DS}}$ & $m_{\mathrm{DS}}$ & $\sigma_{\mathrm{DS}}$ & $Q_{\mathrm{DS}.95}$ \\
 & &  (ns) & (ns) & (ns) & (ns) \\
\hline
 \multirow{2}{*}{3.7 GHz}& LOS  & 20.3 & 20.4 & 4.0 & 26.7 \\
\cline{2-6}
 						 & NLOS & 37.4 & 36.3 & 5.7 & 45.4 \\
\hline
 \multirow{2}{*}{28 GHz} & LOS  & 23.1 & 22.8 & 4.2 & 30.4 \\
\cline{2-6}
						 & NLOS & 33.6 & 35.7 & 5.6 & 41.8\\
\hline\hline
 & & \multicolumn{2}{c|}{$\mu_{lgDS}$} & \multicolumn{2}{c|}{$\sigma_{lgDS}$} \\
\hline
\multirow{2}{*}{3.7 GHz} & LOS  & \multicolumn{2}{c|}{-7.70} & \multicolumn{2}{c|}{0.09} \\
\cline{2-6}
						 & NLOS & \multicolumn{2}{c|}{-7.43} & \multicolumn{2}{c|}{0.07} \\
\hline
3.7 GHz & LOS  & \multicolumn{2}{c|}{-7.49} & \multicolumn{2}{c|}{0.13} \\
\cline{2-6}
\cite{jaeckel2019industrial} & NLOS & \multicolumn{2}{c|}{-7.36} & \multicolumn{2}{c|}{0.11} \\
\hline
\multirow{2}{*}{28 GHz}  & LOS  & \multicolumn{2}{c|}{-7.64} & \multicolumn{2}{c|}{0.08} \\
\cline{2-6}
						 & NLOS & \multicolumn{2}{c|}{-7.48} & \multicolumn{2}{c|}{0.08} \\
\hline
\multirow{2}{*}{3GPP}    & LOS  & \multicolumn{2}{c|}{-7.85} & \multicolumn{2}{c|}{0.15} \\
\cline{2-6}
						 & NLOS & \multicolumn{2}{c|}{-7.72} & \multicolumn{2}{c|}{0.19} \\
\hline
\end{tabular}
\end{table}

\begin{figure}[htb]
\vspace{.1em}
\begin{subfigure}{.48\textwidth}
    \centering
    \includegraphics[width=0.95\textwidth]{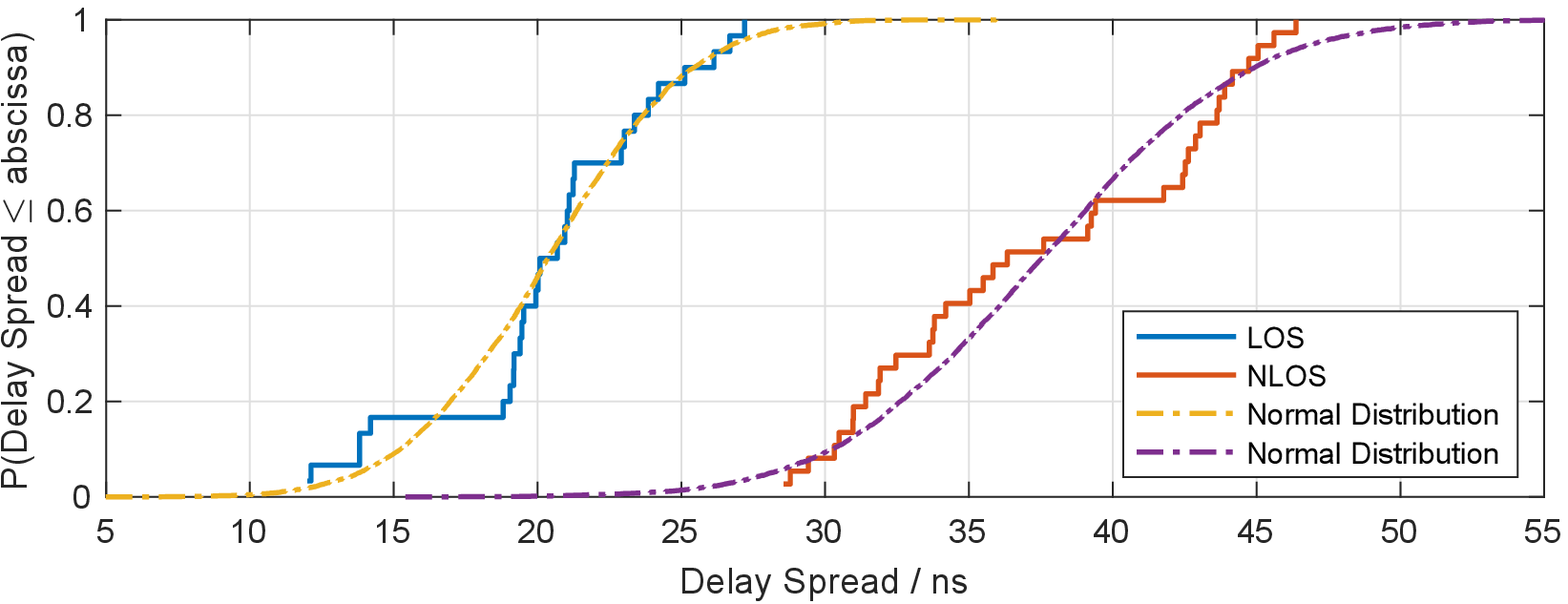}
    \caption{3.7\,GHz}
    \label{fig:ds_37}
\end{subfigure}

\begin{subfigure}{.48\textwidth}
    \centering
    \includegraphics[width=0.95\textwidth]{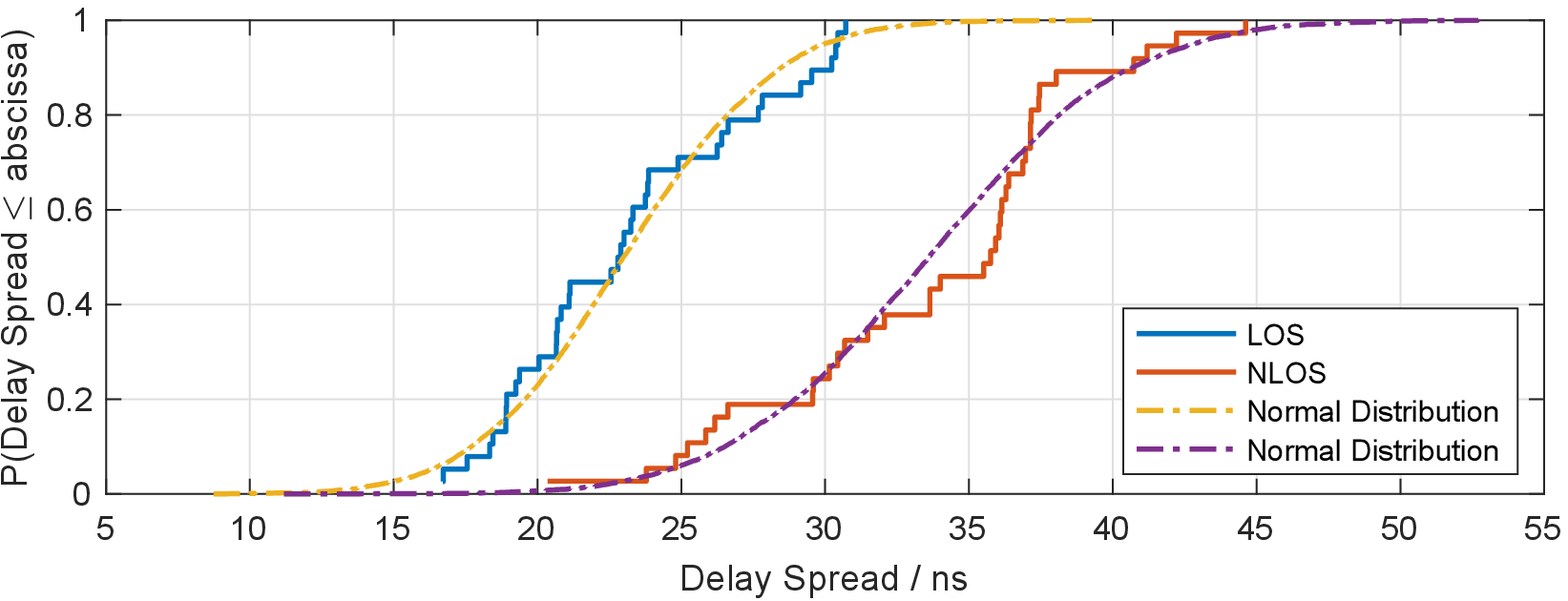}
    \caption{28\,GHz}
    \label{fig:ds_28}
\end{subfigure}
\caption{CDF of RMS delay spread in LOS and NLOS condition; Normal distributions parameterized accordingly to fit the DS CDFs}
\label{fig:ds_lin}
\end{figure}

The mean value of the DS in LOS is 17.1 and 10.5\,ns lower than in NLOS at 3.7 and 28\,GHz, respectively, which can also be seen in the cumulative distribution functions (CDF), displayed in Figure \ref{fig:ds_lin}. Compared to the findings in \cite{schmieder2019directional}, where the mean DS values at 28\,GHz were 53.6\,ns for LOS and 75.3\,ns for NLOS, and to the parameters in \cite{jaeckel2019industrial}, the results in this paper show a much lower delay spread. The 3GPP model parameters however are even smaller, but of the same magnitude as the results in this paper.

\vspace{-2pt}
\subsection{Angular Spread}
\begin{figure}[htb]
    \centering
    \includegraphics[width=0.45\textwidth]{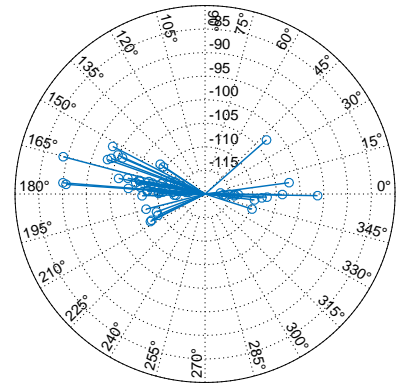}
    \caption{Angular power profile in LOS condition at 28\,GHz}
    \label{fig:apdp_los}
\end{figure}

\begin{figure}[htb]
    \centering
    \includegraphics[width=0.45\textwidth]{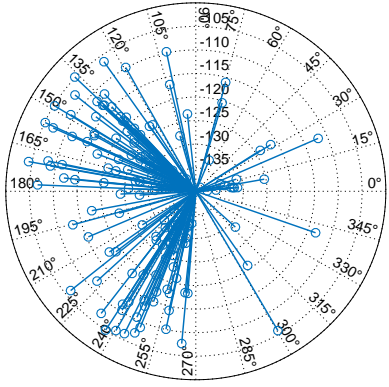}
    \caption{Angular power profile in NLOS condition at 28\,GHz}
    \label{fig:apdp_nlos}
\end{figure}

Direction of arrival (DoA) information was extracted from the CIRs at 28\,GHz in the form of a temporally and spatially resolved list of discrete propagation paths and their powers relative to the total signal received power. This was used to estimate angular power profiles (APP) and the RMS azimuth spread of arrival (ASA). Figures \ref{fig:apdp_los} and \ref{fig:apdp_nlos} show the APPs in LOS and NLOS condition, corresponding to the APDPs in Figure \ref{fig:pdp}. In Figure \ref{fig:apdp_los}, the strong LOS component can be seen impinging from 165\,\textdegree{} together with several strong specular MPCs between 150\,\textdegree{} and 210\,\textdegree{} and between 345\,\textdegree{} and 45\,\textdegree.
Figure \ref{fig:apdp_nlos} shows a much higher spread of MPCs from 105\,\textdegree{} to 270\,\textdegree{} and from 300\,\textdegree{} to 75\,\textdegree{}. 

\begin{figure}[htb]
    \centering
    \includegraphics[width=0.45\textwidth]{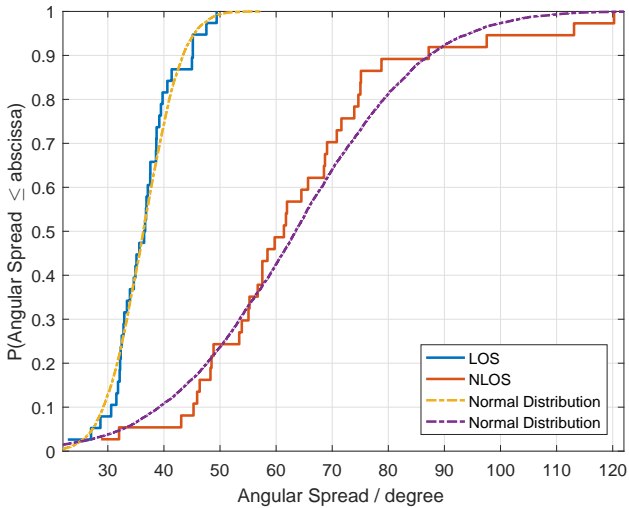}
    \caption{CDF of RMS angular spread in LOS and NLOS condition at 28\,GHz; Normal distributions parameterized accordingly to fit the ASA CDFs}
    \label{fig:as_lin}
\end{figure}

The statistical RMS ASA parameters mean $\mu_{\mathrm{ASA}}$, median $m_{\mathrm{ASA}}$, standard deviation $\sigma_{\mathrm{ASA}}$ and 95\%-quantile $Q_{\mathrm{ASA},95}$ are given in Table \ref{tab:angularspread}. For better comparability with the 3GPP \emph{Indoor Factory} model, the parameters are again given both in linear and logarithmic scaling.
The mean difference between LOS and NLOS AS is 27.2\,\textdegree{} which can also be seen in the CDF displayed in Figure \ref{fig:as_lin}. Comparison with the 3GPP \emph{Indoor Factory} model reveals a good fit and the LOS results in \cite{schmieder2019directional} with $\mu_{\mathrm{ASA,LOS}} = 35.38$\,ns are very similar to the findings in this paper. The NLOS ASA in \cite{schmieder2019directional} with $\mu_{\mathrm{ASA,NLOS}} = 80.01$\,ns is higher than the spread in this paper, but still in a similar range.

\begin{table}[htb]
\renewcommand{\arraystretch}{1.0}
\caption{Statistical parameters of ASA with 3GPP InF \cite{3GPP-TR-38901} for comparison}
\label{tab:angularspread}
\centering
\begin{tabular}{|c|c|c|c|c|}
\hline
%
%
 & $\mu_{\mathrm{ASA}}$ (\degree) & $m_{\mathrm{ASA}}$ (\degree) & $\sigma_{\mathrm{ASA}}$ (\degree)& $Q_{\mathrm{ASA}.95} (\degree)$ \\
\hline
LOS  & 36.4 & 36.5 & 5.5 & 46.6\\
\hline
NLOS & 63.6 & 61.3 & 19.1 & 107.6 \\
\hline%
%
%
\hline%
 & $\mu_{\mathrm{lgASA}}$ & 3GPP $\mu_{\mathrm{lgASA}}$ & $\sigma_{\mathrm{lgASA}}$ & 3GPP $\sigma_{\mathrm{lgASA}}$ \\
\hline
LOS  & 1.55 & 1.52 & 0.07 & 0.4 \\
\hline
NLOS & 1.78 & 1.72 & 0.13 & 0.3 \\
\hline%
\end{tabular}
\end{table}

\vspace{-2pt}
\section{Conclusion}
In this paper, a wideband channel measurement campaign at 3.7 and 28\,GHz in an industrial environment was presented. Based on the captured CIR snapshots, power delay profiles were evaluated that show that the radio channel comprises few specular multipath components and is packed with dense multipath components with a delay of up to 600\,ns. A frequency dependent ABG path loss model was fitted using the results at 3.7 and 28\,GHz in LOS and NLOS condition. Comparison with recent results of other papers and the novel 3GPP TR 38.901 \emph{Indoor Factory} model show that the path loss characteristics are unique and highly scenario dependent. Evaluation of the RMS delay spread show a smaller spread than in similar recent results. They are however of the same magnitude as the 3GPP \emph{Indoor Factory} model parameters. At 28\,GHz, direction-of-arrival information was extracted and angular power profiles and RMS angular spread were evaluated. Comparison with the 3GPP \emph{Indoor Factory} model show that the model agrees well with the estimated RMS angular spread. Overall, the results in this paper show that industrial environments are highly unique and while a first model has been standardized by 3GPP, further evaluation of such scenarios is still relevant, especially in connection with dense or diffuse multipath components.

\section*{Acknowledgment}
This work has been supported by the Governing Mayor of Berlin, Senate Chancellery -- Higher Education and Research, and the European Union -- European Regional Development Fund.





%
\bibliographystyle{IEEEtran}
\bibliography{IEEEabrv,references}

\end{document}